\documentclass[12pt,preprint]{aastex}

\shorttitle{K-vacancy levels in Ne, Mg, Si, S, Ar, and Ca}

\shortauthors{Palmeri et al.}

\begin{document}

\title{Radiative and Auger decay of K-vacancy levels in the
Ne, Mg, Si, S, Ar, and Ca isonuclear sequences}

\author{P. Palmeri,}
\affil{Astrophysique et Spectroscopie, Universit\'e de
Mons-Hainaut, B-7000 Mons, Belgium}
\email{patrick.palmeri@umh.ac.be}

\author{P. Quinet,}
\affil{Astrophysique et Spectroscopie, Universit\'e de
Mons-Hainaut, B-7000 Mons, Belgium and IPNAS, Sart Tilman B15,
Universit\'e de Li\`ege, B-4000 Li\`ege, Belgium}
\email{pascal.quinet@umh.ac.be}

\author{C. Mendoza, M. A. Bautista,}
\affil{Centro de F\'{\i}sica, IVIC, Caracas 1020A, Venezuela}
\email{claudio@ivic.ve; mbautist@ivic.ve}

\author{J. Garc\'{\i}a, and T. R. Kallman}
\affil{NASA Goddard Space Flight Center, Greenbelt, MD 20771}
\email{javier@milkyway.gsfc.nasa.gov; timothy.r.kallman@nasa.gov}


\begin{abstract}
The {\sc hfr} and {\sc autostructure} atomic structure codes are
used to compute complete data sets of level energies, wavelengths,
$A$-values, and radiative and Auger widths for K-vacancy states of
the Ne, Mg, Si, S, Ar, and Ca isonuclear sequences. Ions with
electron number $N>9$ are treated for the first time. Detailed
comparisons with previous measurements and theoretical data for
ions with $N\leq 9$ are carried out in order to estimate reliable
accuracy ratings.
\end{abstract}

\keywords{atomic processes --- atomic data --- line formation
--- X-rays: general}


\section{Introduction}
While X-ray emission is characteristic of hot ($T\geq 10^6$~K)
plasmas, X-ray spectra can also be used to study plasmas of lower
ionization via inner-shell transitions.  Processes which
preferentially remove inner-shell electrons, such as
photoionization or collisions with suprathermal particles, can
give rise to emission of inner-shell fluorescence lines or K-edge
absorption in astrophysical spectra.  These features are observed
from several important classes of astrophysical sources, being
potential diagnostics of their conditions: ionization, gas
motions, abundances, and inner-shell ionization rate.  The grating
instruments on the {\em Chandra} and {\em XMM-Newton} space
observatories have provided high signal-to-noise spectra of these
features, with typical energy resolution of
$\varepsilon/\Delta\varepsilon\simeq 500-1000$.  Exploitation of
these diagnostics has been limited by the accuracy of the atomic
data for these transitions (energies, cross sections, and
lifetimes). This paper is part of a large effort to calculate as
many of these data as possible with reliable accuracy using
available computational methods.  Previous papers in this series
have focussed on the data for iron and oxygen; in this and
subsequent papers we concentrate on medium-$Z$ elements, namely
Ne, Mg, Si, S, Ar, and Ca.

In the X-ray spectra obtained with {\em Chandra} and {\em
XMM-Newton}, K lines and edges from medium-$Z$ elements have been
observed from compact objects in which photoionization is thought
to be the dominant ionization mechanism, and from low ionization
material which is seen along the line of sight to various bright
sources. The active galactic nucleus NGC 3783 observed by
\citet{kas02} shows K-absorption lines of H- and He-like Ne, Mg,
Si, S, possibly Ar and Ca, and from lower stages of ionization of
Si and S. They are found to be blueshifted and asymmetric, and no
correlation is obtained between the velocity shifts and the
ionization potential of the ions.  Absorption from low and medium
stages of Si and S are also seen in the spectra from AGNs
MCG-6-30-15 \citep{Lee01} and IRAS 13349+2438 \citep{Holc07}, and
from the X-ray binary Cyg X-1 \citep{Chan07}. In emission these
lines are seen in the spectra from Vela X-1 by \citet{Wata06} and
4U1700-37 by \citet{Boro03}. Furthermore, it is likely that future
observations, which provide high resolution X-ray spectra from
spatially extended sources, will reveal inner-shell features in
supernova remnants and stellar coronae.  In these sources, such
features may indicate existence of non-thermal particles or
non-equilibrium ionization, and their analysis requires accurate data
for the wavelengths and other atomic quantities associated with
the K lines from medium-$Z$ elements.

Following the work by \citet{pal02,pal03a,pal03b},
\citet{bau03,bau04}, \citet{men04}, and \citet{kal04} on the K
lines of Fe and by \citet{gar05} on the K-shell photoabsorption of
O ions, we report new atomic data for K-vacancy levels in the Ne,
Mg, Si, S, Ar, and Ca isonuclear sequences. Prime objectives are
to improve the atomic database of the {\sc xstar} modelling code
\citep{bau01} and to prepare ionic targets (configuration
expansions and orbitals) for the lengthy computations of the
photoabsorption cross sections where both radiative and Auger
dampings are key effects. In this respect, available atomic
structure data sets---namely K-vacancy level energies,
wavelengths, $A$-values, and Auger widths---for first-row ions
with electron number $2\leq N\leq 10$ are far from complete while
for the second row ($11\leq N\leq 20$) they hardly exist.

Previous work on the K-shell structure of medium-$Z$ elements
includes that by \citet{fae94} on the satellites of the He-like
resonance line in species with atomic number $12\leq Z\leq 16$ and
electron number $4\leq N\leq 9$. They have measured wavelengths in
a CO$_2$ laser produced plasma and computed wavelengths,
$A$-values, and Auger rates with the {\sc mz} code \citep{vai78,
vai80}. \citet{bie00} have measured wavelengths for Ar ions with
$3\leq N\leq 9$ in a plasma focus discharge, providing also
theoretical wavelengths, $A$-values, and Auger rates calculated
with both the {\sc hfr} atomic structure code \citep{cow81} and
the multiconfiguration Dirac--Fock (MCDF) {\sc yoda} code of
\citet{hag87}. K-vacancy level energies, wavelengths, $A$-values,
and Auger and radiative widths have been computed for ions of the
beryllium  ($6\leq Z\leq 26$), boron ($6\leq Z\leq 54$), and
carbon ($6\leq Z\leq 54$) isoelectronic sequences with the MCDF
method by \citet{che85}, \citet{che88}, and \citet{che97},
respectively. Inner-shell excitation energies and autoionization
rates for ions with $6\leq Z\leq 54$ and $6\leq N\leq 9$ have been
computed by \citet{saf99} using the $1/Z$ perturbation theory
method. \citet{beh02} have calculated with the relativistic
multiconfiguration {\sc hullac} code \citep{bar01} wavelengths and
oscillator strengths for the 1s-$n$p transitions ($n\leq 3$) in
ions of Ne, Mg, Al, Si, S, Ar, and Ca with $2\leq N\leq 9$.
\citet{des03} have produced a comprehensive compilation of both
measured and theoretical transition energies for K lines and edges
in elements with $10\leq Z\leq 100$. \citet{gor03} have audited
the fluorescence database by \citet{kaa93} which is widely used in
modelling codes, in particular their scaling procedures along
isoelectronic sequences. The have found serious flaws which appear
to compromise the application of this database in plasma
modelling.

The outline of the present report is as follows. The numerical
methods are briefly described in Section~2 while an analysis of
the results based on comparisons with previous experimental and
theoretical values is carried out in Section~3. The two
supplementary electronic tables are explained in Section~4 while
some conclusions are finally discussed in Section~5.


\section{Numerical methods}
The numerical approach has been fully described by \citet{bau03}.
The atomic data are computed with the structure codes {\sc hfr}
\citep{cow81} and {\sc autostructure} \citep{eis74, bad86, bad97}.
Wavefunctions are calculated with the Breit--Pauli relativistic
corrections
\begin{equation}
  \label{hbp}
  H_{\rm bp} = H_{\rm nr} + H_{\rm 1b} + H_{\rm 2b}
\end{equation}
where $H_{\rm nr}$ is the usual non-relativistic Hamiltonian. The
one-body relativistic operators
\begin{equation}
   \label{h1b}
   H_{\rm 1b} = \sum_{n=1}^{N} {f_n({\rm mass}) + f_n({\rm d}) + f_n({\rm so})}
\end{equation}
represent the spin--orbit interaction, $f_n({\rm so})$, and the
non-fine-structure mass variation, $f_n({\rm mass})$, and the
one-body Darwin correction, $f_n({\rm d})$. The two-body Breit
operators are given by
\begin{equation}
   \label{h2b}
   H_{\rm 2b} = \sum_{n<m} g_{nm}({\rm so}) + g_{nm}({\rm ss}) + g_{nm}({\rm css})
   + g_{nm}({\rm d}) + g_{nm}({\rm oo})
\end{equation}
where the fine-structure terms are $g_{nm}(\rm so)$
(spin--other-orbit and mutual spin-orbit) and $g_{nm}(\rm ss)$
(spin--spin), and the non-fine-structure counterparts are
$g_{nm}(\rm css)$ (spin--spin contact), $g_{nm}(\rm d)$ (two-body
Darwin), and $g_{nm}(\rm oo)$ (orbit--orbit). Core relaxation
effects are studied with {\sc autostructure} by comparing ion
representations where all the electron configurations have a
common basis of orthogonal orbitals, to be referred to hereafter
as approximation AS1, with one where each configuration has its
own basis, approximation AS2. {\sc hfr} computes energies,
$A$-values, and Auger rates with bases of non-orthogonal orbitals
obtained by optimizing the energy of each configuration, and
neglects the part of the Breit interaction (\ref{h2b}) that cannot
be reduced to a one-body operator. This data set is to be labelled
HFR1.


\section{Results}

We have carried out detailed comparisons with previous data in
order to obtain accuracy estimates and detect weak points. In this
respect, \citet{pal06} have already presented the outcome for the
S isonuclear sequence. Their comparison with the experimental
K-level energies for S~{\sc xv} and S~{\sc xiv} results in
differences not larger than 1~eV for HFR1 and 2~eV for AS2, but
for the only level reported for S~{\sc i} ($[{\rm 1s}]{\rm 3p}^5\
^3{\rm P}^{\rm o}_2$), differences of 3.3~eV and 3.6~eV are
respectively found. The level of agreement between HFR1 and AS2
reported by Palmeri et al. for the S sequence is $\sim$1 eV for
level energies, $\sim$2 m\AA\ for wavelengths, and $\sim$10\% for
radiative and Auger rates greater than $10^{13}$~s$^{-1}$. They
cannot explain, however, large discrepancies with the MCDF data by
\citet{che97} for C-like S~{\sc xi}. Since a similar degree of
discord with Chen et al. was encountered by \citet{pal03a} for
Fe~{\sc xxi}, we emphasize here comparisons with the experimental
and theoretical data sets for Ar ions by \citet{bie00},
particularly those computed with the MCDF method, as they will
provide more solid evidence on the questionable accuracy of the
data by Chen et al.

\subsection{Energy levels}

With the exception of the He-, Li-, and Be-like species, there is
a general lack of experimental K-vacancy level energies for the
isonuclear sequences under study here. This severely limits the
possibilities of fine-tuning the numerical methods in order to
increase data accuracy and to provide reliable rankings. For this
reason, we mainly base our approach on computations with the {\sc
hfr} code as it attains marginally more reliable {\em ab initio}
energies due to the use of non-orthogonal basis sets for each
electronic configuration.

\begin{figure}
\epsscale{.80} \plotone{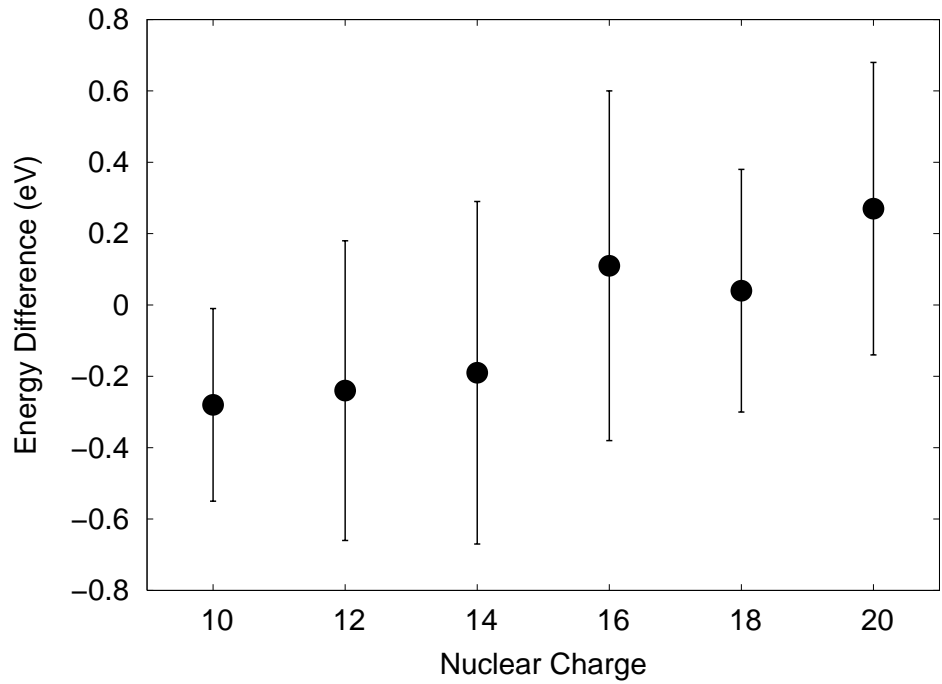}
\caption{\label{nist} Average differences between the HFR1 energies and those in the
NIST database V3.1.2 \citep{ral07} for the K-vacancy levels in the He and Li
isoelectronic sequences. Error bars indicate the standard deviation. For nuclear charge
$Z=18$, the experimental level energies for the Li-like ion have been excluded as they
are believed to be incorrect.}
\end{figure}

In Fig.~\ref{nist} we show the average energy differences between
the values listed in the NIST database V3.1.2 \citep{ral07} and
HFR1 for the K-vacancy levels in He- and Li-like ions. Differences
are never larger than 1~eV if those for Li-like Ar~{\sc xvi} are
excluded as the NIST K-vacancy level energies for this ion are
believed to be grossly in error. As listed in Table~\ref{e-ar},
the energy difference for the ${\rm 1s}{\rm 2s}^2\ ^2{\rm
S}_{1/2}$ level in Ar~{\sc xvi} is 0.1~eV while those for the rest
are $\sim$756 eV. The experimental inaccuracies for this ion are
further established with the wavelength measurement by
\citet{bei02} for the ${\rm 1s2s2p}\ ^4{\rm P}^{\rm
o}_{5/2}\rightarrow {\rm 1s}^2{\rm 2s}\ ^2{\rm S}_{1/2}$
dipole-forbidden transition. They use high-resolution spectroscopy
of a low-temperature tokamak plasma to obtain an energy of
3090.25$\pm$0.12 eV for the upper level which is in remarkable
agreement with the HFR1 value of 3090.1 eV in Table~\ref{e-ar}.
For Be-like Mg and Si the agreement between HFR1 and NIST is well
within 2~eV.

Theoretical K-vacancy level energies for ions with $18\leq Z\leq
26$ and $6\leq N\leq 9$ are listed by \citet{saf99} which allow a
comparison with HFR1 for species with $18\leq Z\leq 20$. It is
found that energy differences are within 3~eV except for the eight
levels in both Ar and Ca shown in Table~\ref{e-SS}. They mainly
belong to the O-like type, and the HFR1 values are always larger
by as much as 6~eV.


\subsection{Wavelengths}

Since theoretical wavelengths do not reach spectroscopic accuracy,
statistical comparisons with measurements are essential inasmuch
as they enable small empirical adjustments to make the computed
values astrophysically useful. It is worth pointing out that, due
to the lack of experimental data for ionic species of the second
row, this procedure is only currently possible for systems with
electron number $N\leq 9$.

\begin{figure}
\epsscale{.80} \plotone{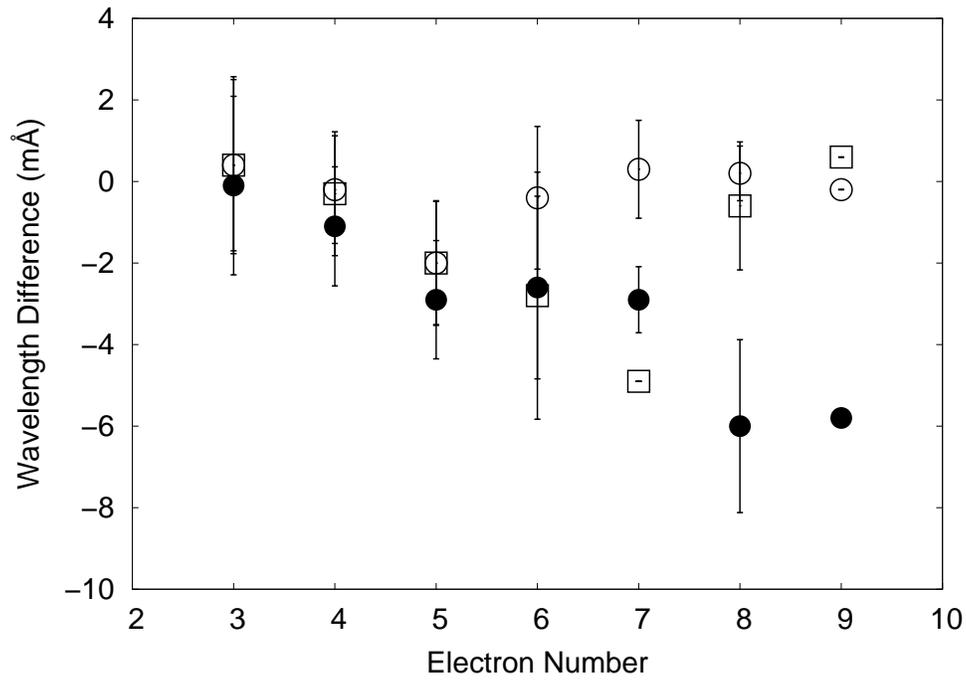}
\caption{\label{wl} Average wavelength difference $\overline{\Delta \lambda_{\rm
e}}$ (m\AA) as a function of the electron number for ions in the Ar isonuclear sequence.
$\overline{\Delta \lambda_{\rm e}}$ is determined with respect to the spectroscopic
values of \citet{bie00}. Filled circles: HFR1 values. Circles: HFR2 results by
\citet{bie00}. Squares: MCDF1 results by \citet{bie00}. A large error bar for the
MCDF1 value at $N=7$ has been removed for clarity.}
\end{figure}

Experimental wavelengths for Ar ions with $3\leq N\leq 9$ have
been reported by \citet{bie00} as well as theoretical values
computed with the {\sc hfr} and {\sc yoda} codes. The two latter
data sets are to be hereafter referred to as the HFR2 and MCDF1,
respectively. As shown in Fig.~\ref{wl}, a statistical comparison
with the measured values can be carried out where the average
theory--experiment wavelength difference, $\overline{\Delta
\lambda}_{\rm e}$, is plotted as a function of the ion electron
number $N$. It may be seen that theoretical wavelengths appear to
be always shorter than the measured values. In the case of our
HFR1 data, $|\overline{\Delta \lambda}_{\rm e}|$ increases with
$N$ from under 1~m\AA\ for $N=3$ to around 6~m\AA\ for $8\leq
N\leq 9$. A similar behavior is displayed by the MCDF1 values
except for $8\leq N\leq 9$ where $|\overline{\Delta \lambda_{\rm
e}}|< 1$~m\AA; this seems to indicate that the transition energies
in these two ions have been adjusted with the experimental values.
This is certainly also the case with the HFR2 results for $3\leq
N\leq 9$ where $|\overline{\Delta \lambda_{\rm e}}| < 1$~m\AA.
However, as shown in Fig.~\ref{wl}, the standard deviations for
all three theoretical data sets are around 2--3~m\AA\ which gives
an indication of the level of accuracy that can be attained
theoretically.

For the computed MCDF wavelengths listed by \citet{che88} and
\citet{che97} for B- and C-like Ar, $|\overline{\Delta
\lambda_{\rm e}}|=2\pm 2$~m\AA\ and $|\overline{\Delta
\lambda_{\rm e}}|=4\pm 6$~m\AA, respectively. The somewhat large
standard deviation in the C-like system is caused by transitions
involving heavily mixed K-vacancy states, in particular ${\rm
1s(^2S)2s2p^4(^2D)\ ^1D_2}$, but the level of agreement is in
general similar to that discussed above for HFR1.

Differences between the HFR1 wavelengths and the measurements
listed by \citet{fae94} for ions with $12\leq Z\leq 16$ and $4\leq
N\leq 9$ display a similar behavior to that found in the Ar ions:
for each isonuclear sequence, $|\overline{\Delta \lambda_{\rm
e}}|$ increases with $N$. Also, as expected, it decreases with $Z$
along the isoelectronic sequence; for instance, in the case of
$N=9$, $|\overline{\Delta \lambda_{\rm e}}|\approx 18$~m\AA\ for
$Z=14$ while  $|\overline{\Delta \lambda_{\rm e}}|\approx
10$~m\AA\ for $Z=16$. On the other hand, we have found three
transitions in N-like ions ($N=7$) with unusually large wavelength
differences which are listed in Table~\ref{w-faenov}.

As a result of the comparisons with the spectroscopic data, the HFR1
wavelengths for systems with $3\leq N\leq 9$ have been empirically
shifted with $-\overline{\Delta \lambda_{\rm e}}$.

\citet{beh02} list wavelengths computed with the {\sc hullac} code
for 68 K-transitions arising from the ground level of ions with
$2\leq N\leq 9$. In general, the agreement with HFR1 is better
than 10~m\AA\ and progressively improves with $Z$ up to 3~m\AA\
for $Z=20$. As depicted in Table~\ref{w-hullac}, there are however
several transitions with noticeable differences, specially for
ions with $Z\leq 14$ and $6\leq N\leq 9$ where the longer {\sc
hullac} wavelengths are probably more accurate than HFR1.
Surprisingly large differences (10 m\AA) are found between {\sc
hullac} and measurements for transitions in Li-like Ne for which
the measured values agree with HFR1 to within 3~m\AA\ (see
Table~\ref{w-hullac}).

HFR1 K$\alpha$, K$\beta$, and edge transition energies for singly
ionized species with $10\leq Z\leq 20$ are compared in
Table~\ref{ktrans} with the compilation by \citet{des03} of
experimental and theoretical X-ray transition energies. The HFR1
edge transition energy is estimated using the experimental
ionization potential of the neutral and the HFR1 position for the
lowest K-vacancy level of the singly ionized stage. The agreement
with their theoretical and the experimental K$\alpha$ transition
energies is within 1.5 eV, but discrepancies as large as 7~eV are
found with the experimental K$\beta$ values and with the edge
energies ($\sim$10 eV) for $12\leq Z\leq 16$. The soundness of
this comparison may be limited by experimental difficulties in
assigning weak features and by solid-state effects.


\subsection{$A$-values}
Present $A$-values are computed with approximations AS1, AS2, and
HFR1. As mentioned in Section~2, in AS2 and HFR1 non-orthogonal
orbital bases are used which then account for orbital relaxation
in the radiative decay process. It has been found that the 1s
orbital is particularly different in valence and K-vacancy states.
Since orthogonal bases are used in AS1, a comparison of AS1 and
AS2 would give an indication of the importance of this effect.

\begin{figure}
\epsscale{.80} \plotone{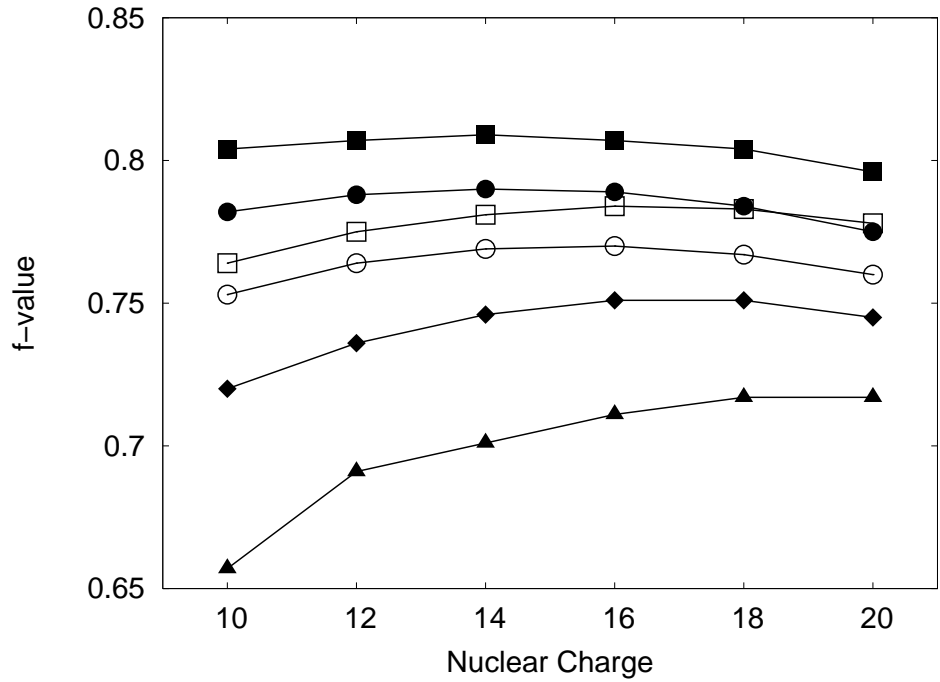}
\caption{\label{f-he} Absorption $f$-value for the ${\rm 1s}^2\ ^1{\rm S}_0\rightarrow {\rm 1s2p}\
^1{\rm P}^{\rm o}_1$ resonance transition in the He isoelectronic
sequence. Filled squares: HFR1. Filled cirles: AS2. Circles: AS1.
Squares: HFR, single configuration. Filled diamonds: values by
\cite{dra79} calculated in a unified relativistic theory. Filled
triangles: data computed with the {\sc hullac} code by \cite{beh02}.}
\end{figure}

The radiative data listed by \citet{beh02} for K transitions,
namely $f$- and $A$-values, are consistently (10$\pm 5$)\% below
HFR1. To study this situation, we first look at the absorption
$f$-value for the ${\rm 1s}^2\ ^1{\rm S}_0\rightarrow {\rm 1s2p}\
^1{\rm P}^{\rm o}_1$ resonance transition along the He
isoelectronic sequence (see Fig.~\ref{f-he}). It may be seen that
the $f$-values by Behar \& Netzer are about 6\% below those by
\cite{dra79} which are usually taken as the reference. On the
other hand, HFR1 is 9\% higher and so is AS2 (6\%). As shown in
Fig.~\ref{f-he}, if the HFR1 $f$-values are recomputed in a
single-configuration approximation, the differences with Drake are
reduced to 6\% and AS1 also does somewhat better (3\%). The latter
result suggests that perhaps taking into account core relaxation
effects does not necessarily lead to improved radiative data. This
hypothesis is confirmed by a comparison of available lifetime
measurements in He- and Li-like ions with AS1 and AS2 (see
Table~\ref{lifet}). The computed lifetimes for these $n=2$ levels
are sensitive to the wavefuntions because most of them involve
optically forbidden decay channels. Although the agreement of both
AS1 and AS2 with experiment is within approximately 10\%, it is
not clear which is the most accurate. As a conclusion, we are only
confident that the accuracy of the HFR1 $A$-values is at around
the 15\% level.

This suggested level of accuracy is further supported by a
comparison of the HFR1 $A$-values greater than $10^{13}$~s$^{-1}$
in Ar ions with those computed by \citet{bie00} in the HFR2 and
MCDF1 approximations. In spite of their fine tuning of the
transition matrix elements with the experimental energy levels in
HFR2, the general agreement is well within 15\% except for the few
problematic transitions listed in Table~\ref{aval}. It may be seen
that for transitions in species with $N=5$ and $N=7$, HFR1 and
HFR2 are in reasonable agreement in contrast to MCDF1. For $N=6$,
the scatter is large which is surely due to strong admixture of
the upper K-vacancy levels produced by the spin-orbit interaction.

\begin{figure}
\epsscale{.80} \plotone{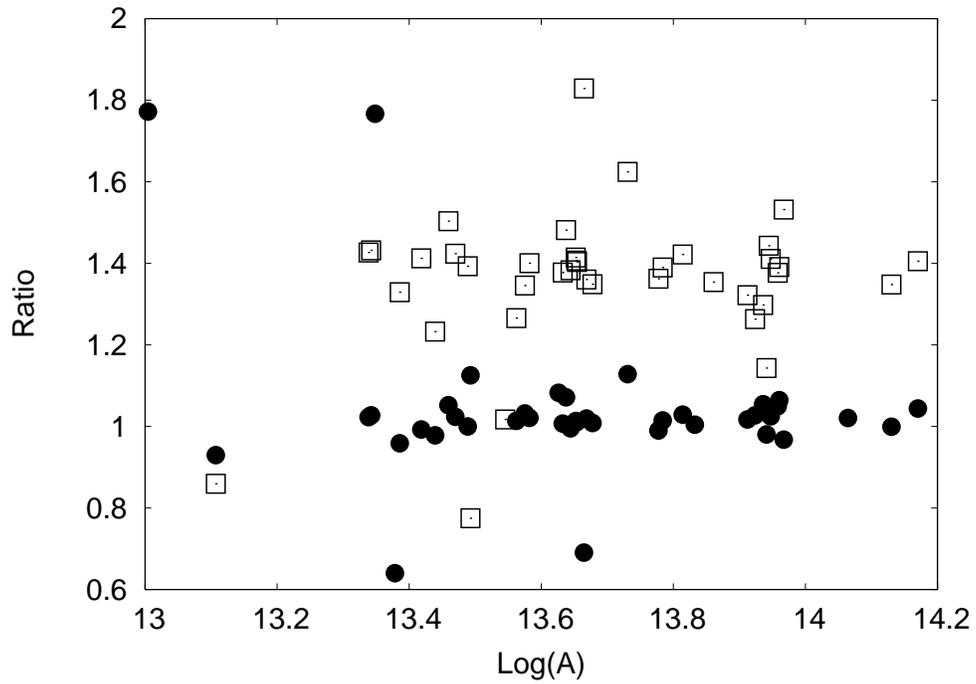}
\caption{\label{chen-A} Comparison of the present HFR1 $A$-values
for K transitions in C-like Ar with two independent MCDF calculations. Squares:
\cite{che97}. Filled circles: MCDF1 $A$-values by \cite{bie00}. It is found
that the data by \cite{che97} are on average larger by 37\% with respect to HFR1.}
\end{figure}

$A$-values calculated by \citet{che88} for B-like Ar using the
MCDF method agree within 20\% with HFR1 although on average they
are $\sim$10\% smaller. However, we find that those computed for
C-like Ar by \citet{che97}, as shown in Fig.~\ref{chen-A}, are
37\% higher. We also include in this plot the MCDF1 $A$-values by
\citet{bie00} which are within 15\% of HFR1 if the problematic
values of Table~\ref{aval} are excluded. Moreover, there are
several transitions by Chen et al. that show differences larger
than a factor of 2 with HFR1 which have not been taken into
account in this comparison.

The comparison of the HFR1 $A$-values with those calculated by
\citet{fae94} using the {\sc mz} method for ions with $12\leq
Z\leq 16$ and $4\leq N\leq 9$ shows a wide scatter: some
transitions agree to better than 10\% while large discrepancies
are found for others.


\subsection{Radiative and Auger widths}
Auger widths of K-vacancy levels are determined by including all
the decay channels $C\epsilon l$, where $C$ are all the $n=2$
valence configurations of the $(N-1)$-electron daughter ion and
$l\leq 4$.

\citet{bie00} have computed with {\sc hfr} and the MCDF {\sc yoda}
code Auger widths for K-vacancy states in Ar ions with $3\leq
N\leq 9$ (data sets referred to as HFR2 and MCDF1, respectively).
A reasonable agreement (within 20\%) with HFR1 is found except for
the two levels in the Li-like system ($N=3$) and the five levels
in the Be-like system ($N=4$) shown in Table~\ref{auger}. Also the
MCDF1 K-vacancy levels in the N-like ion ($N=7$) are on average
around 40\% larger. The discrepant HFR2 results for the Li-like
ion are attributed to typos since a recalculation results in data
very similar to the HFR1 values.

\begin{figure}
\epsscale{.80} \plotone{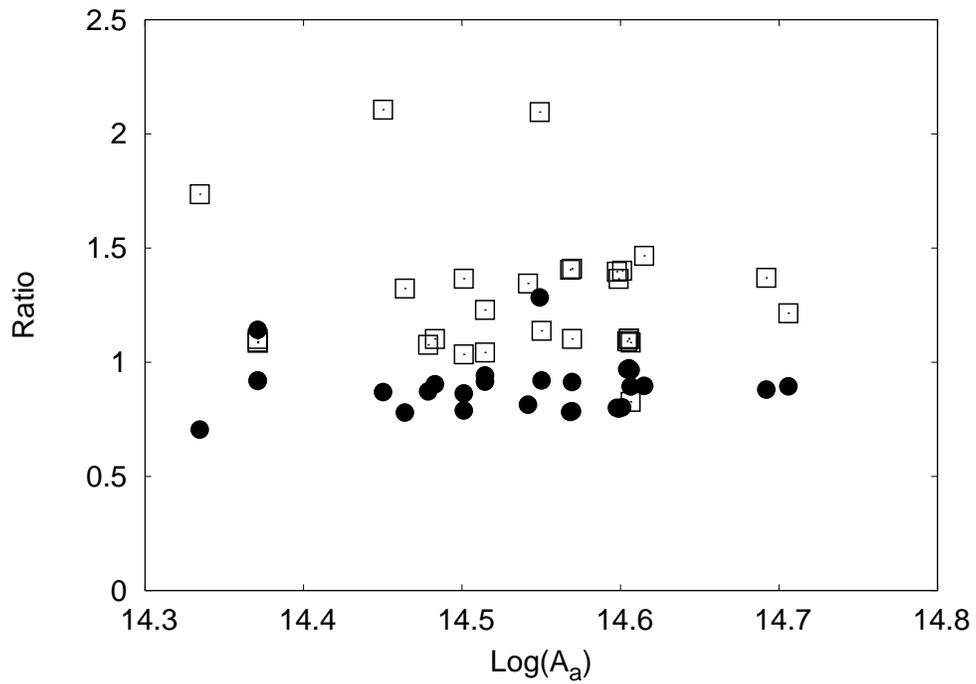}
\caption{\label{chen-Au} Comparison of the present HFR1 Auger
widths for K-vacancy levels in C-like Ar with two independent MCDF
calculations. Squares: \cite{che97}. Filled circles: MCDF1
Auger widths by \cite{bie00}. It is found that the data by
\cite{che97} are on average larger by 30\%.}
\end{figure}

MCDF radiative and Auger widths have also been computed for Be-,
B-, and C-like Ar by \citet{che85}, \citet{che88}, and
\citet{che97}, respectively. For the Be- and B-like systems, they
agree with HFR1 to within 10\% except for the small Auger widths
($A_{\rm a}< 10^{12}$~s$^{-1}$)  of the ${\rm 1s2s2p^2\ ^5P_j}$
levels in Be-like Ar. For the C-like system, on the other hand,
radiative and Auger widths by Chen et al. are on average larger
than HFR1 by 40\% and 30\%, respectively, with some differences
being as large as a factor of 2. Such a poor accord can be
appreciated in Fig.~\ref{chen-Au} where the MCDF1 data by
\citet{bie00} is also shown. The latter agrees with HFR1 to within
20\%.

The total K-widths quoted by \citet{beh02} follow a similar trend
to that found for their $A$-values: on average they are found to
be about 15\% below HFR1, although for some levels large
discrepancies are encountered (see Table~\ref{a-hullac}). For such
levels, it is shown that the agreement between HFR1 and AS1 is
around 5\%.

The Auger rates for ions with atomic number $12\leq Z\leq 16$ and
electron number $4\leq N\leq 9$ computed with the {\sc mz} code by
\citet{fae94} are found to be in complete disagreement with HFR1.

In their study of the fluorescence database by \citet{kaa93},
\citet{gor03} have computed with {\sc autostructure}
configuration-averaged fluorescence yields for the K-vacancy
configurations ${\rm 1s2s}^2{\rm 2p}$ in Be-like ions and ${\rm
1s2s}^2{\rm 2p}^6$ in F-like ions. They also list yields estimated
from the MCDF data of \citet{che85} and from the widths computed
with {\sc hullac} by \citet{beh02}, the latter having been revised
due to missing Auger channels and misquoted widths. In the
comparison presented in Table~\ref{yields}, it may be seen that
our HFR1 fluorescence yields agree with those by \citet{gor03} to
within 10\% except for the Ne F-like ion (20\%). Differences may
be due to the non-orthogonal orbital bases and neglect of two-body
effects in the {\sc hfr} code. The agreement between HFR1 and
\citet{beh02} for the Be-like configuration is within 6\% but is
poorer for the F-like (20\%), and that with \citet{che85} for the
Be-like ion is better than 15\%.

\section{Supplementary electronic tables}
Computed energy levels, wavelengths, $A$-values, $gf$-values, and
radiative and Auger widths for ions in the Ne, Mg, Si, S, Ar, and
Ca isonuclear sequences can be accessed from the online
machine-readable
Tables~11\footnote{https://hartree.ivic.ve/tempfiles/tab11.txt}--12\footnote{https://hartree.ivic.ve/tempfiles/tab12.txt}.
The printed samples show data for ions of the Ne isonuclear with
electron number $N\leq 3$. It may be seen that in Table~11 levels
are identified with the vector $(Z,N,i,2S+1,L,2J,{\rm Conf})$
where $Z$ is the atomic number, $N$ the electron number, $2S+1$
the spin multiplicity, $L$ the total orbital angular momentum
quantum number, $J$ the total angular momentum quantum number, and
Conf the level configuration assignment. For each level, the
spectroscopic energy (when available), the computed HFR1 energy,
and its radiative width ($A_{\rm r}$) are listed. For K-vacancy
levels, the Auger width ($A_{\rm a}$) and fluorescence yield are
additionally given. In Table~12, transitions are identified with
the vector $(Z,N,k,i)$ where $k$ and $i$ are the upper and lower
level indices, respectively, tabulating its computed wavelength,
$A$-value, and $gf$-value.


\section{Summary and conclusions}
Extensive data sets containing energy levels, wavelengths,
$A$-values and radiative and Auger widths for K-vacancy levels in
the Ne, Mg, Si, S, Ar, and Ca isonuclear sequences have been
computed with the atomic structure codes {\sc hfr} and {\sc
autostructure} which include relativistic corrections. For ionic
species of the second row with electron numbers $11\leq N\leq 20$,
it is the first time that such data become available. For
first-row ions ($2\leq N\leq 10$), detailed comparisons have been
carried out with available measurements and theoretical values
which have brought forth the consistency and accuracy of the
present data sets.

Comparisons of the present HFR1 K-vacancy level energies with
those listed in the NIST database and with recent measurements
support an accuracy rating for ions with electron number $N\leq 4$
of better than 2~eV. Furthermore, we have shown that the NIST
energies for K vacancy levels in Li-like Ar are incorrect.
Comparisons of the HFR1 wavelengths with the spectroscopic values
reported for ions with $N\leq 9$ by \citet{fae94} and
\citet{bie00} show that the theoretical wavelengths are
consistently shorter. Moreover, along an isonuclear sequence, the
average theory--experiment wavelength difference
$|\overline{\Delta \lambda}_{\rm e}|$ increases with $N$. This
behavior enables empirical corrections to be made to the HFR1
wavelengths in ions with $N\leq 9$ that are bound to improve
accuracy to around 2--3~m\AA. Due to a complete absence of
spectroscopic measurements for K transitions in ions of the second
row, this fine-tuning procedure cannot be extended at present to
species with $N>9$ whose wavelengths therefore remain uncorrected.

The $A$-values and Auger widths for the Ar isonuclear sequence
previously computed by \citet{bie00} with {\sc hfr} and the MCDF
{\sc yoda} code have given us the opportunity to benchmark the
accuracy of the present data sets. This has been useful to sort
out the large discrepancies reported by \citet{pal03a} for Fe~{\sc
xxi} and \citet{pal06} for S~{\sc xi} with the MCDF data of
\citet{che97}. Similar discrepancies are found with their data for
Ar~{\sc xiii} which, in the light of the good agreement of HFR1
with Bi\'emont et al., are certainly not due to the more formal
relativistic representation of the MCDF method. Significant
discords of HFR1 with the Auger widths reported by \citet{fae94}
for the Mg, Si, and S isonuclear sequences and with those by
\citet{beh02} for levels in the B and C isoelectronic sequences
are difficult to explain. Accuracy ratings of the present
$A$-values and Auger widths greater than $10^{13}$~s$^{-1}$ are
confidently assigned to 15\% and 20\%, respectively, except for
special cases; e.g. transitions involving K-vacancy levels in the
carbon isoelectronic sequence where strong admixture causes severe
departures.

The present radiative and Auger widths will be used in the
computations of the K-shell photoionization cross sections of
these medium-$Z$ ions which are required in XSTAR for the
modelling of interesting spectral features. Future work will
involve extension of the present approach to the Ni isonuclear
sequence.


\acknowledgments
MAB acknowledges partial support from FONACIT,
Venezuela, under contract No. S1-20011000912.  This work
was  funded in part by the NASA Astronomy and
Physics Research and Analysis Program.   Partial support for CM was
provided  by a travel grant from IVIC and
the FNRS of Belgium.



\clearpage
\begin{deluxetable}{lll}
\tablecolumns{3}
\tablewidth{0pc} \tablecaption{\label{e-ar} K-vacancy level energies (eV) in Ar~{\sc xvi}}
\tablehead{\colhead{Level} & \colhead{Expt$^a$} & \colhead{HFR1} \\}
\startdata
${\rm 1s}{\rm 2s}^2\ ^2{\rm S}_{1/2}$                                       & 3079.0 & 3079.1 \\
${\rm 1s}(^2{\rm S}){\rm 2s2p}(^3{\rm P}^{\rm o})\ ^4{\rm P}^{\rm o}_{1/2}$ & 3842.3 & 3086.7 \\
${\rm 1s}(^2{\rm S}){\rm 2s2p}(^3{\rm P}^{\rm o})\ ^4{\rm P}^{\rm o}_{3/2}$ & 3843.8 & 3087.9 \\
${\rm 1s}(^2{\rm S}){\rm 2s2p}(^3{\rm P}^{\rm o})\ ^4{\rm P}^{\rm o}_{5/2}$ & 3847.1 & 3090.1 \\
${\rm 1s}(^2{\rm S}){\rm 2p}^2(^3{\rm P})\ ^4{\rm P}_{1/2}$                 & 3876.2 & 3121.6 \\
${\rm 1s}(^2{\rm S}){\rm 2p}^2(^3{\rm P})\ ^4{\rm P}_{3/2}$                 & 3878.6 & 3123.0 \\
${\rm 1s}(^2{\rm S}){\rm 2p}^2(^3{\rm P})\ ^4{\rm P}_{5/2}$                 & 3881.0 & 3124.9 \\
${\rm 1s}(^2{\rm S}){\rm 2s2p}(^1{\rm P}^{\rm o})\ ^2{\rm P}^{\rm o}_{1/2}$ & 3884.8 & 3124.9 \\
${\rm 1s}(^2{\rm S}){\rm 2s2p}(^1{\rm P}^{\rm o})\ ^2{\rm P}^{\rm o}_{3/2}$ & 3885.9 & 3125.8 \\
${\rm 1s}(^2{\rm S}){\rm 2p}^2(^1{\rm D})\ ^2{\rm D}_{3/2}$                 & 3896.6 & 3139.8 \\
${\rm 1s}(^2{\rm S}){\rm 2p}^2(^1{\rm D})\ ^2{\rm D}_{5/2}$                 & 3897.2 & 3140.4 \\
\enddata
\tablenotetext{a}{NIST database V3.1.2 \citep{ral07}, T. Shirai et
al. (1999, unpublished)}
\end{deluxetable}


\clearpage
\begin{deluxetable}{llrrrrr}
\tablecolumns{7}
\tablewidth{0pc} \tablecaption{\label{e-SS}K-vacancy level energies (eV) in Ar and Ca}
\tablehead{\colhead{} & \colhead{} & \multicolumn{2}{c}{$Z=18$} & \colhead{} & \multicolumn{2}{c}{$Z=20$} \\
\cline{3-4}\cline{6-7}\\
\colhead{$N$} & \colhead{Level} & \colhead{HFR1} & \colhead{SS$^a$}
& \colhead{} &
   \colhead{HFR1} & \colhead{SS$^a$}\\}
\startdata
6 & ${\rm 1s2p}^5\ ^1{\rm P}^{\rm o}_1$           & 3165.4 & 3162.0 && 3935.5 & 3932.0 \\
7 & ${\rm 1s2p}^6\ ^2{\rm S}_{1/2}$               & 3157.8 & 3154.1 && 3925.7 & 3921.9 \\
8 & ${\rm 1s2s}^2{\rm 2p}^5\ ^3{\rm P}^{\rm o}_2$ & 2992.3 & 2988.4 && 3734.8 & 3730.6 \\
8 & ${\rm 1s2s}^2{\rm 2p}^5\ ^3{\rm P}^{\rm o}_1$ & 2994.0 & 2990.4 && 3737.5 & 3733.8 \\
8 & ${\rm 1s2s}^2{\rm 2p}^5\ ^3{\rm P}^{\rm o}_0$ & 2995.2 & 2991.9 && 3739.4 & 3736.3 \\
8 & ${\rm 1s2s}^2{\rm 2p}^5\ ^1{\rm P}^{\rm o}_1$ & 3004.7 & 3000.5 && 3749.9 & 3745.4 \\
8 & ${\rm 1s2s2p}^6\ ^3{\rm S}_1$                 & 3059.8 & 3054.6 && 3813.3 & 3808.0 \\
8 & ${\rm 1s2s2p}^6\ ^1{\rm S}_0$                 & 3076.9 & 3071.2 && 3832.9 & 3827.3 \\
\enddata
\tablenotetext{a}{\cite{saf99}}
\end{deluxetable}


\clearpage
\begin{deluxetable}{rrllrrr}
\tablecolumns{7} \tablewidth{0pc}
\tablecaption{\label{w-faenov}HFR1 and experimental wavelengths
(\AA) with large differences} \tablehead{\colhead{$Z$} &
\colhead{$N$} & \colhead{Upper level} & \colhead{Lower level} &
\colhead{HFR1} & \colhead{Expt$^a$} & \colhead{MZ$^a$}\\}
\startdata
14 & 7 & ${\rm 1s2p}^6\ ^2{\rm S}_{1/2}$                         & ${\rm 1s^22s^22p^3}\ ^2{\rm P^o_{3/2}}$ & 6.6590 & 7.0156 & 7.0161 \\
16 & 7 & ${\rm 1s(^2S)2s2p^5(^3P^{\rm o})}\ ^4{\rm P}^{\rm o}_{1/2}$ & ${\rm 1s^22s2p^4}\ ^2{\rm D}_{3/2}$ & 5.3141 & 5.2804 & 5.2797 \\
   & 7 & ${\rm 1s2p}^6\ ^2{\rm S}_{1/2}$                         & ${\rm 1s^22s^22p^3}\ ^2{\rm P^o_{3/2}}$ & 5.0434 & 5.2863 & 5.2864 \\
\enddata
\tablenotetext{a}{Experiment and calculation ({\sc mz} code) by \citet{fae94}}
\end{deluxetable}


\clearpage
\begin{deluxetable}{rrllrrr}
\tablecolumns{7}
\tablewidth{0pc} \tablecaption{\label{w-hullac}HFR1 and HULLAC
wavelengths (\AA) with large differences} \tablehead{\colhead{$Z$}
& \colhead{$N$} & \colhead{Upper level} & \colhead{Lower level} &
\colhead{HFR1} & \colhead{HULLAC$^a$} & \colhead{Expt}\\}
\startdata
10 & 3 & ${\rm 1s}{\rm 2s2p}\ ^2{\rm P}^{\rm o}_{1/2}$ & ${\rm 1s}^2{\rm 2s}\ ^2{\rm S}_{1/2}$                     & 13.658 & 13.648 & 13.655(3)$^b$ \\
   & 3 & ${\rm 1s}{\rm 2s2p}\ ^2{\rm P}^{\rm o}_{3/2}$ & ${\rm 1s}^2{\rm 2s}\ ^2{\rm S}_{1/2}$                     & 13.656 & 13.646 & 13.655(3)$^b$ \\
   & 4 & ${\rm 1s2s}^2{\rm 2p}\ ^1{\rm P}^{\rm o}_1$   & ${\rm 1s}^2{\rm 2s}^2\ ^1{\rm S}_0$                       & 13.824 & 13.814 & \\
   & 6 & ${\rm 1s2s}^2{\rm 2p}^3\ ^3{\rm D}^{\rm o}_1$ & ${\rm 1s}^2{\rm 2s}^2{\rm 2p}^2\ ^3{\rm P}_0$             & 14.225 & 14.239 & \\
   & 8 & ${\rm 1s2s}^2{\rm 2p}^5\ ^3{\rm P}^{\rm o}_2$ & ${\rm 1s}^2{\rm 2s}^2{\rm 2p}^4\ ^3{\rm P}_2$             & 14.490 & 14.526 & \\
   & 9 & ${\rm 1s2s}^2{\rm 2p}^6\ ^2{\rm S}_{1/2}$     & ${\rm 1s}^2{\rm 2s}^2{\rm 2p}^5\ ^2{\rm P}^{\rm o}_{3/2}$ & 14.596 & 14.631 & \\
12 & 6 & ${\rm 1s2s}^2{\rm 2p}^3\ ^3{\rm D}^{\rm o}_1$ & ${\rm 1s}^2{\rm 2s}^2{\rm 2p}^2\ ^3{\rm P}_0$             &  9.622 &  9.631 & \\
   & 8 & ${\rm 1s2s}^2{\rm 2p}^5\ ^3{\rm P}^{\rm o}_2$ & ${\rm 1s}^2{\rm 2s}^2{\rm 2p}^4\ ^3{\rm P}_2$             &  9.797 &  9.816 & \\
   & 9 & ${\rm 1s2s}^2{\rm 2p}^6\ ^2{\rm S}_{1/2}$     & ${\rm 1s}^2{\rm 2s}^2{\rm 2p}^5\ ^2{\rm P}^{\rm o}_{3/2}$ &  9.875 &  9.895 & 9.9129$^c$ \\
14 & 8 & ${\rm 1s2s}^2{\rm 2p}^5\ ^3{\rm P}^{\rm o}_2$ & ${\rm 1s}^2{\rm 2s}^2{\rm 2p}^4\ ^3{\rm P}_2$             &  7.052 &  7.063 & 7.0665$^c$ \\
   & 9 & ${\rm 1s2s}^2{\rm 2p}^6\ ^2{\rm S}_{1/2}$     & ${\rm 1s}^2{\rm 2s}^2{\rm 2p}^5\ ^2{\rm P}^{\rm o}_{3/2}$ &  7.107 &  7.119 & 7.1244$^c$ \\
\enddata
\tablenotetext{a}{\citet{beh02}} \tablenotetext{b}{NIST database V3.1.2 \citep{ral07}, \cite{kra06}}
\tablenotetext{c}{\citet{fae94}}
\end{deluxetable}


\clearpage
\begin{deluxetable}{llllll}
\tablecolumns{7}
\tablewidth{0pc}
\tablecaption{\label{ktrans}Transition energies (eV)} \tablehead{\colhead{$Z$} & \colhead{K$\alpha_1$} &
\colhead{K$\alpha_2$} &
   \colhead{K$\beta_1$} & \colhead{K$\beta_3$} & \colhead{Edge} \\}
\startdata
10 & 849.5        & 849.4        &             &             & 871.0       \\
   & 849.2        & 849.1        &             &             & 870.7       \\
   & 848.61(26)   & 848.61(26)   & 857.89(44)  & 857.89(44)  & 870.23(18)  \\
12 & 1254.6       & 1254.6       & 1308.4      & 1308.4      & 1311.5      \\
   & 1254.4       & 1254.1       &             &             & 1312.3      \\
   & 1253.688(11) & 1253.437(13) & 1302.20(40) & 1302.20(40) & 1303.33(27) \\
14 & 1741.2       & 1741.2       & 1841.8      & 1842.8      & 1849.2      \\
   & 1741.2       & 1739.7       &             & 1841.8      & 1850.3      \\
   & 1739.985(19) & 1739.394(34) & 1835.96(40) & 1835.96(40) & 1839.13(37) \\
16 & 2309.2       & 2306.8       & 2468.7      & 2470.2      & 2481.9      \\
   & 2308.8       & 2307.0       & 2467.5      & 2469.7      & 2481.7      \\
   & 2307.885(34) & 2306.700(38) & 2464.07(14) &             & 2471.63(70) \\
18 & 2958.7       & 2956.6       & 3192.7      & 3192.5      & 3208.4      \\
   & 2957.9       & 2955.9       & 3191.5      & 3191.3      & 3207.4      \\
   & 2957.682(16) & 2955.566(16) & 3190.49(24) & 3190.49(24) & 3206.14(54) \\
20 & 3692.4       & 3688.8       & 4015.3      & 4015.0      & 4049.9      \\
   & 3691.0       & 3687.6       & 4014.7      & 4014.3      & 4049.4      \\
   & 3691.719(49) & 3688.128(49) & 4012.76(38) & 4012.76(38) & 4050.48(30) \\
\enddata
\tablecomments{For each element, the first row lists HFR1 data, the second and third rows
respectively give computed and measured values by \citet{des03}.}
\end{deluxetable}


\clearpage
\begin{deluxetable}{llllll}
\tablecolumns{6} \tabletypesize{\scriptsize} \tablewidth{0pc}
\tablecaption{\label{lifet} Lifetimes for He- and Li-like ions}
\tablehead{\colhead{Level} & \colhead{$Z$}& \colhead{$N$} & \multicolumn{3}{c}{$\tau$ (ns)}\\
\cline{4-6}\\
\colhead{} & \colhead{} & \colhead{} & \colhead{AS1} & \colhead{AS2} & \colhead{Expt}\\}
\startdata
${\rm 1s2s}\ ^3{\rm S}_1$         & 10 & 2 & 9.10$+$4 & 1.01$+$5 & 9.17(4)$+$4$^a$   \\
                                  & 12 & 2 & 1.37$+$4 & 1.50$+$4 & 1.361(49)$+$4$^b$ \\
                                  & 16 & 2 & 7.04$+$2 & 7.55$+$2 & 7.03(4)$+$2$^c$   \\
                                  & 18 & 2 & 2.11$+$2 & 2.24$+$2 & 2.03(12)$+$2$^d$  \\
${\rm 1s2p}\ ^3{\rm P}^{\rm o}_0$ & 18 & 2 & 5.11     & 4.60     & 4.87(44)$^e$      \\
${\rm 1s2p}\ ^3{\rm P}^{\rm o}_1$ & 12 & 2 & 3.26$-$2 & 3.07$-$2 & 2.90(15)$-$2$^f$  \\
                                  & 14 & 2 & 7.08$-$3 & 6.73$-$3 & 6.45(30)$-$3$^f$, 6.35(33)$-$3$^g$ \\
                                  & 16 & 2 & 1.93$-$3 & 1.85$-$3 & 1.57(18)$-$3$^g$  \\
${\rm 1s2p}\ ^3{\rm P}^{\rm o}_2$ & 16 & 2 & 2.80     & 2.60     & 2.5(2)$^h$        \\
                                  & 18 & 2 & 1.58     & 1.50     & 1.62(8)$^e$       \\
${\rm 1s2p}\ ^1{\rm P}^{\rm o}_1$ & 14 & 2 & 2.59$-$5 & 2.52$-$5 & 2.9(10)$-$5$^i$   \\
\tableline
${\rm 1s2s2p}\ ^4{\rm P}^{\rm o}_{5/2}$ & 10 & 3 & 1.09$+$1 & 1.01$+$1 & 1.04(15)$+$1$^j$ \\
                                        & 14 & 3 & 2.19     & 2.07     & 2.1(1)$^k$  \\
                                        & 16 & 3 & 1.14     & 1.09     & 1.1(2)$^h$  \\
                                        & 18 & 3 & 6.33$-$1 & 6.08$-$1 & 5.94(16)$-$1$^l$  \\
${\rm 1s2p}^2\ ^4{\rm P}_{1/2}$         & 10 & 3 & 5.46$-$1 & 4.75$-$1 & 5.3(5)$-$1$^m$ \\
                                        & 12 & 3 & 1.56$-$1 & 1.40$-$1 & 1.74(17)$-$1$^n$ \\
${\rm 1s2p}^2\ ^4{\rm P}_{3/2}$         & 10 & 3 & 3.90$-$1 & 3.46$-$1 & 4.0(4)$-$1$^m$ \\
                                        & 12 & 3 & 9.38$-$2 & 8.35$-$2 & 9.4(12)$-$2$^n$ \\
${\rm 1s2p}^2\ ^4{\rm P}_{5/2}$         & 12 & 3 & 8.80$-$3 & 7.45$-$3 & 9(4)$-$3$^n$ \\
\enddata
\tablenotetext{a}{\cite{tra99}}
\tablenotetext{b}{\cite{ste95}}
\tablenotetext{c}{\cite{cre06}}
\tablenotetext{d}{\cite{hub87}}
\tablenotetext{e}{\cite{dav77}}
\tablenotetext{f}{\cite{arm81}}
\tablenotetext{g}{\cite{var76a}}
\tablenotetext{h}{\cite{coc74}}
\tablenotetext{i}{\cite{var76b}}
\tablenotetext{j}{\cite{gro75}}
\tablenotetext{k}{\cite{has75}}
\tablenotetext{l}{\cite{doh79}}
\tablenotetext{m}{\cite{kny85}}
\tablenotetext{n}{\cite{hel85}}
\tablecomments{$a\pm b\equiv a\times 10^{\pm b}$.}
\end{deluxetable}

\clearpage
\begin{deluxetable}{rlllrrr}
\tablecolumns{7}
\tablewidth{0pc} \tablecaption{\label{aval} Transitions in Ar ions with questionable $A$-values (s$^{-1}$)}
\tablehead{\colhead{$N$}& \colhead{Upper level} & \colhead{Lower level} & \colhead{HFR1} &
\colhead{HFR2$^a$} & \colhead{MCDF1$^a$}\\}
\startdata
5 & ${\rm 1s(^2S)}{\rm 2s2p}^3{\rm (^3D^{\rm o})}\ ^4{\rm D}^{\rm o}_{3/2}$ & ${\rm 1s}^2{\rm 2s2p}^2\ ^4{\rm P}_{3/2}$           & 1.49$+$13 & 1.48$+$13 & 1.08$+$13 \\
  & ${\rm 1s(^2S)}{\rm 2s2p}^3{\rm (^1D^{\rm o})}\ ^2{\rm D}^{\rm o}_{3/2}$ & ${\rm 1s}^2{\rm 2s2p}^2\ ^2{\rm D}_{5/2}$           & 2.36$+$13 & 2.20$+$13 & 1.77$+$13 \\
  & ${\rm 1s(^2S)}{\rm 2s2p}^3{\rm (^1P^{\rm o})}\ ^2{\rm P}^{\rm o}_{1/2}$ & ${\rm 1s}^2{\rm 2s2p}^2\ ^2{\rm S}_{1/2}$           & 2.62$+$13 & 2.97$+$13 & 1.72$+$13 \\
  & ${\rm 1s}{\rm 2p}^4\ ^2{\rm D}_{3/2}$                                   & ${\rm 1s}^2{\rm 2p}^3\ ^2{\rm P}^{\rm o}_{1/2}$     & 1.02$+$13 & 9.00$+$12 & 6.80$+$12 \\
6 & ${\rm 1s2s}^2{\rm 2p}^3\ ^3{\rm S}^{\rm o}_1$                           & ${\rm 1s}^2{\rm 2s}^2{\rm 2p}^2\ ^3{\rm P}_2$       & 3.11$+$13 & 2.33$+$13 & 3.50$+$13 \\
  & ${\rm 1s(^2S)2s}{\rm 2p}^4{\rm (^2D)}\ ^3{\rm D}_{2}$                   & ${\rm 1s}^2{\rm 2s}{\rm 2p}^3\ ^3{\rm D}^{\rm o}_2$ & 7.26$+$13 & 1.28$+$13 & 2.00$+$11 \\
  & ${\rm 1s(^2S)2s}{\rm 2p}^4{\rm (^2D)}\ ^3{\rm D}_{2}$                   & ${\rm 1s}^2{\rm 2s}{\rm 2p}^3\ ^3{\rm D}^{\rm o}_3$ & 3.45$+$13 & 6.29$+$13 & 9.60$+$13 \\
  & ${\rm 1s(^2S)2s}{\rm 2p}^4{\rm (^2D)}\ ^3{\rm D}_{2}$                   & ${\rm 1s}^2{\rm 2s}{\rm 2p}^3\ ^3{\rm P}^{\rm o}_1$ & 2.66$+$11 & 3.52$+$13 & 3.51$+$13 \\
  & ${\rm 1s(^2S)2s}{\rm 2p}^4{\rm (^2D)}\ ^3{\rm D}_{2}$                   & ${\rm 1s}^2{\rm 2s}{\rm 2p}^3\ ^3{\rm P}^{\rm o}_2$ & 3.29$+$13 & 4.70$+$12 & 1.74$+$13 \\
  & ${\rm 1s(^2S)2s}{\rm 2p}^4{\rm (^2D)}\ ^3{\rm D}_{1}$                   & ${\rm 1s}^2{\rm 2s}{\rm 2p}^3\ ^3{\rm P}^{\rm o}_2$ & 8.07$+$12 & 1.12$+$13 & 5.80$+$12 \\
  & ${\rm 1s(^2S)2s}{\rm 2p}^4{\rm (^4P)}\ ^3{\rm P}_{2}$                   & ${\rm 1s}^2{\rm 2s}{\rm 2p}^3\ ^3{\rm D}^{\rm o}_3$ & 6.94$+$13 & 4.58$+$13 & 1.50$+$13 \\
  & ${\rm 1s(^2S)2s}{\rm 2p}^4{\rm (^4P)}\ ^3{\rm P}_{2}$                   & ${\rm 1s}^2{\rm 2s}{\rm 2p}^3\ ^3{\rm P}^{\rm o}_2$ & 7.57$+$12 & 3.89$+$13 & 2.50$+$13 \\
  & ${\rm 1s(^2S)2s}{\rm 2p}^4{\rm (^2D)}\ ^1{\rm D}_{2}$                   & ${\rm 1s}^2{\rm 2s}{\rm 2p}^3\ ^1{\rm P}^{\rm o}_1$ & 1.01$+$13 & 1.45$+$13 & 1.79$+$13 \\
  & ${\rm 1s(^2S)2s}{\rm 2p}^4{\rm (^2P)}\ ^3{\rm P}_{2}$                   & ${\rm 1s}^2{\rm 2s}{\rm 2p}^3\ ^1{\rm D}^{\rm o}_2$ & 2.23$+$13 & 1.39$+$13 & 3.94$+$13 \\
  & ${\rm 1s(^2S)2s}{\rm 2p}^4{\rm (^2P)}\ ^3{\rm P}_{2}$                   & ${\rm 1s}^2{\rm 2s}{\rm 2p}^3\ ^3{\rm S}^{\rm o}_1$ & 2.39$+$13 & 3.22$+$13 & 1.53$+$13 \\
  & ${\rm 1s(^2S)2s}{\rm 2p}^4{\rm (^2P)}\ ^3{\rm P}_{2}$                   & ${\rm 1s}^2{\rm 2s}{\rm 2p}^3\ ^1{\rm P}^{\rm o}_1$ & 1.29$+$13 & 9.30$+$12 & 5.50$+$12 \\
7 & ${\rm 1s(^2S)2s2p}^5{\rm (^3P^{\rm o})}\ ^2{\rm P}^{\rm o}_{3/2}$       & ${\rm 1s}^2{\rm 2s2p}^4\ ^2{\rm S}_{1/2}$           & 2.67$+$13 & 2.72$+$13 & 5.30$+$12 \\
  & ${\rm 1s(^2S)2s2p}^5{\rm (^3P^{\rm o})}\ ^2{\rm P}^{\rm o}_{1/2}$       & ${\rm 1s}^2{\rm 2s2p}^4\ ^2{\rm D}_{3/2}$           & 1.39$+$14 & 1.41$+$14 & 2.94$+$13 \\
  & ${\rm 1s(^2S)2s2p}^5{\rm (^1P^{\rm o})}\ ^2{\rm P}^{\rm o}_{1/2}$       & ${\rm 1s}^2{\rm 2s2p}^4\ ^2{\rm P}_{3/2}$           & 2.58$+$13 & 2.61$+$13 & 1.35$+$14 \\
\enddata
\tablenotetext{a}{HFR2 and MCDF1 results computed by \citet{bie00}}
\tablecomments{$a\pm b\equiv a\times 10^{\pm b}$.}
\end{deluxetable}


\clearpage
\begin{deluxetable}{llllll}
\tablecolumns{6}
\tablewidth{0pc} \tablecaption{\label{auger} Questionable Auger widths (s$^{-1}$) for Ar ions}
\tablehead{\colhead{$N$} & \colhead{Level} & \colhead{HFR1} & \colhead{AS1} & \colhead{HFR2$^a$} &
\colhead{MCDF1$^a$}\\} \startdata
 3 & ${\rm 1s(^2S)2p^2(^3P)}\ ^2{\rm P}_{1/2}$     & 1.34$+$11 & 1.19$+$11 & 1.70$+$12 & 2.00$+$11 \\
   & ${\rm 1s(^2S)2p^2(^1S)}\ ^2{\rm S}_{1/2}$     & 2.41$+$13 & 2.15$+$13 & 6.39$+$13 & 2.21$+$13 \\
 4 & ${\rm 1s2s}^2{\rm 2p}\ ^1{\rm P}^{\rm o}_{1}$ & 1.14$+$14 & 1.20$+$14 & 5.69$+$13 & 1.08$+$14 \\
   & ${\rm 1s(^2S)2s2p^2(^4P)}\ ^3{\rm P}_1$       & 9.75$+$13 & 9.62$+$13 & 1.10$+$14 & 5.23$+$13 \\
   & ${\rm 1s(^2S)2s2p^2(^2D)}\ ^3{\rm D}_1$       & 8.63$+$13 & 9.57$+$13 & 7.53$+$13 & 5.23$+$13 \\
   & ${\rm 1s(^2S)2s2p^2(^4P)}\ ^3{\rm P}_2$       & 4.23$+$13 & 3.73$+$13 & 4.96$+$13 & 5.41$+$13 \\
   & ${\rm 1s2p}^3\ ^3{\rm S}^{\rm o}_{1}$         & 5.06$+$12 & 6.23$+$12 & 1.90$+$13 & 4.50$+$12 \\
\enddata
\tablenotetext{a}{HFR2 and MCDF1 results computed by \citet{bie00}}
\tablecomments{$a\pm b\equiv a\times 10^{\pm b}$.}
\end{deluxetable}


\clearpage
\begin{deluxetable}{llllll}
\tablecolumns{6}
\tablewidth{0pc} \tablecaption{\label{a-hullac}Questionable HULLAC
K-vacancy level widths (s$^{-1}$)}
\tablehead{\colhead{$Z$}&\colhead{$N$}&\colhead{Level}& \colhead{HFR1} &
\colhead{AS1} & \colhead{HULLAC$^a$}\\} \startdata
 10 & 5 & ${\rm 1s2s}^2{\rm 2p}^2\ ^2{\rm P}_{1/2}$     & 1.03$+$14 & 1.08$+$14 & 3.55$+$13 \\
    & 6 & ${\rm 1s2s}^2{\rm 2p}^3\ ^3{\rm S}^{\rm o}_1$ & 1.06$+$14 & 1.06$+$14 & 5.76$+$13 \\
    & 8 & ${\rm 1s2s}^2{\rm 2p}^5\ ^3{\rm P}^{\rm o}_2$ & 3.71$+$14 & 3.71$+$14 & 2.55$+$14 \\
    & 9 & ${\rm 1s2s}^2{\rm 2p}^6\ ^2{\rm S}_{1/2}$     & 3.92$+$14 & 3.88$+$14 & 5.80$+$12 \\
 12 & 5 & ${\rm 1s2s}^2{\rm 2p}^2\ ^2{\rm P}_{1/2}$     & 1.20$+$14 & 1.33$+$14 & 5.08$+$13 \\
    & 6 & ${\rm 1s2s}^2{\rm 2p}^3\ ^3{\rm S}^{\rm o}_1$ & 1.44$+$14 & 1.48$+$14 & 7.95$+$13 \\
 14 & 5 & ${\rm 1s2s}^2{\rm 2p}^2\ ^2{\rm P}_{1/2}$     & 1.63$+$14 & 1.67$+$14 & 7.58$+$13 \\
    & 6 & ${\rm 1s2s}^2{\rm 2p}^3\ ^3{\rm S}^{\rm o}_1$ & 2.05$+$14 & 2.23$+$14 & 1.16$+$14 \\
 16 & 5 & ${\rm 1s2s}^2{\rm 2p}^2\ ^2{\rm P}_{1/2}$     & 2.11$+$14 & 2.13$+$14 & 1.14$+$14 \\
    & 6 & ${\rm 1s2s}^2{\rm 2p}^3\ ^3{\rm S}^{\rm o}_1$ & 2.80$+$14 & 2.95$+$14 & 1.70$+$14 \\
 18 & 5 & ${\rm 1s2s}^2{\rm 2p}^2\ ^2{\rm P}_{1/2}$     & 2.76$+$14 & 2.76$+$14 & 1.69$+$14 \\
    & 6 & ${\rm 1s2s}^2{\rm 2p}^3\ ^3{\rm S}^{\rm o}_1$ & 3.57$+$14 & 3.63$+$14 & 2.46$+$14 \\
 20 & 5 & ${\rm 1s2s}^2{\rm 2p}^2\ ^2{\rm P}_{1/2}$     & 3.61$+$14 & 3.59$+$14 & 2.47$+$14 \\
    & 6 & ${\rm 1s2s}^2{\rm 2p}^3\ ^3{\rm S}^{\rm o}_1$ & 4.51$+$14 & 4.45$+$14 & 3.46$+$14 \\
\enddata
\tablenotetext{a}{Widths computed with the {\sc hullac} code by
\cite{beh02}} \tablecomments{The level width is given by the sum of the radiative and Auger widths.
Also, $a\pm b\equiv a\times 10^{\pm b}$.}
\end{deluxetable}

\clearpage
\begin{deluxetable}{rrlrrrr}
\tablecolumns{7}
\tablewidth{0pc} \tablecaption{\label{yields} Configuration-averaged fluorescence yields}
\tablehead{\colhead{$Z$} & \colhead{$N$} & \colhead{Configuration} & \colhead{HFR1} &
\colhead{AS3$^a$} & \colhead{HULLAC$^b$} & \colhead{MCDF$^c$}\\}
\startdata
10 & 4 & ${\rm 1s2s}^2{\rm 2p}$   & 0.0223 & 0.0201 & 0.0209 & 0.0191 \\
12 &   &                          & 0.0423 & 0.0393 & 0.0414 & 0.0377 \\
14 &   &                          & 0.0691 & 0.0658 & 0.0685 &        \\
16 &   &                          & 0.1008 & 0.0974 & 0.0984 &        \\
18 &   &                          & 0.1340 & 0.1309 & 0.1273 & 0.1237 \\
20 &   &                          & 0.1673 & 0.1646 & 0.1569 &        \\
10 & 9 & ${\rm 1s2s}^2{\rm 2p}^6$ & 0.0185 & 0.0147 & 0.0215 &        \\
12 &   &                          & 0.0332 & 0.0298 & 0.0380 &        \\
14 &   &                          & 0.0564 & 0.0528 & 0.0630 &        \\
16 &   &                          & 0.0895 & 0.0855 & 0.0983 &        \\
18 &   &                          & 0.1343 & 0.1286 & 0.1443 &        \\
20 &   &                          & 0.1867 & 0.1818 & 0.2001 &        \\
\enddata
\tablenotetext{a}{Computed with {\sc autostructure} by \citet{gor03} including relativistic corrections}
\tablenotetext{b}{Estimated by \citet{gor03} from the widths computed with {\sc hullac} by
\citet{beh02} after some revision}
\tablenotetext{c}{Calculated by \citet{gor03} from the MCDF data
by \citet{che85}}
\end{deluxetable}


\clearpage
\begin{deluxetable}{rrrrrrlrrrrr}
\tablecolumns{12} \tabletypesize{\scriptsize} \tablewidth{0pc}
\tablecaption{\label{levels} Valence and K-vacancy levels for the Ne, Mg, Si, S, Ar, and Ca isonuclear sequences}
\tablehead{\colhead{$Z$} & \colhead{$N$} & \colhead{$i$} & \colhead{$2S+1$} & \colhead{$L$} & \colhead{$2J$} &
\colhead{Level} & \colhead{$E$(NIST)} & \colhead{$E$(HFR1)} & \colhead{$A_{\rm r}(i)$} &
\colhead{$A_{\rm a}(i)$} &  \colhead{Yield}\\
\colhead{} & \colhead{} & \colhead{} & \colhead{} & \colhead{} & \colhead{} & \colhead{} & \colhead{eV} &
\colhead{eV} & \colhead{s$^{-1}$} & \colhead{s$^{-1}$} & \colhead{}\\}
\startdata
10 & 1 &  1 &  2 & 0 & 1 & 1s 2S1/2               &    0.0000 &    0.0000 &          &               &        \\
10 & 1 &  2 &  2 & 1 & 1 & 2p 2P1/2               & 1021.4970 & 1021.4975 &          &               &        \\
10 & 1 &  3 &  2 & 0 & 1 & 2s 2S1/2               & 1021.5180 & 1021.5177 &          &               &        \\
10 & 1 &  4 &  2 & 1 & 3 & 2p 2P3/2               & 1021.9530 & 1021.9528 &          &               &        \\
10 & 2 &  1 &  1 & 0 & 0 & 1s2 1S0                &    0.0000 &    0.0000 &          &               &        \\
10 & 2 &  2 &  3 & 0 & 2 & 1s2s 3S1               &  905.0775 &  904.4935 &          &               &        \\
10 & 2 &  3 &  3 & 1 & 0 & 1s2p 3Po0              &  914.7805 &  914.1721 & 1.03E+08 &               &        \\
10 & 2 &  4 &  3 & 1 & 2 & 1s2p 3Po1              &  914.8177 &  914.2734 & 3.78E+09 &               &        \\
10 & 2 &  5 &  3 & 1 & 4 & 1s2p 3Po2              &  915.0099 &  914.4848 & 1.14E+08 &               &        \\
10 & 2 &  6 &  1 & 0 & 0 & 1s2s 1S0               &  915.3360 &  915.2111 & 1.54E+02 &               &        \\
10 & 2 &  7 &  1 & 1 & 2 & 1s2p 1Po1              &  922.0163 &  921.8277 & 9.88E+12 &               &        \\
10 & 3 &  1 &  2 & 0 & 1 & 1s22s 2S1/2            &    0.0000 &    0.0000 &          &               &        \\
10 & 3 &  2 &  2 & 1 & 1 & 1s22p 2Po1/2           &   15.8888 &   15.8681 & 5.50E+08 &               &        \\
10 & 3 &  3 &  2 & 1 & 3 & 1s22p 2Po3/2           &   16.0933 &   16.0709 & 5.72E+08 &               &        \\
10 & 3 &  4 &  2 & 0 & 1 & 1s2s2 2S1/2            &           &  891.0215 & 5.31E+11 &      1.14E+14 & 0.0046 \\
10 & 3 &  5 &  4 & 1 & 1 & 1s(2S)2s2p(3Po) 4Po1/2 &           &  895.1820 & 2.68E+08 & $<$\ 5.00E+08 & 1.0000 \\
10 & 3 &  6 &  4 & 1 & 3 & 1s(2S)2s2p(3Po) 4Po3/2 &           &  895.2770 & 6.86E+08 &      1.00E+09 & 0.4069 \\
10 & 3 &  7 &  4 & 1 & 5 & 1s(2S)2s2p(3Po) 4Po5/2 &           &  895.4381 & 5.03E+02 &      9.59E+07 & 0.0000 \\
10 & 3 &  8 &  2 & 1 & 1 & 1s(2S)2s2p(3Po) 2Po1/2 &  908.0482 &  907.7819 & 7.98E+12 &      7.73E+12 & 0.5080 \\
10 & 3 &  9 &  2 & 1 & 3 & 1s(2S)2s2p(3Po) 2Po3/2 &  908.0482 &  907.8962 & 8.13E+12 &      6.38E+12 & 0.5603 \\
10 & 3 & 10 &  4 & 1 & 1 & 1s(2S)2p2(3P) 4P1/2    &           &  912.3807 & 1.76E+09 & $<$\ 5.00E+08 & 1.0000 \\
10 & 3 & 11 &  4 & 1 & 3 & 1s(2S)2p2(3P) 4P3/2    &           &  912.4757 & 1.98E+09 &      6.00E+09 & 0.2481 \\
10 & 3 & 12 &  4 & 1 & 5 & 1s(2S)2p2(3P) 4P5/2    &           &  912.6306 & 2.39E+09 &      3.70E+10 & 0.0607 \\
10 & 3 & 13 &  2 & 1 & 1 & 1s(2S)2s2p(1Po) 2Po1/2 &           &  914.3583 & 9.06E+11 &      7.57E+13 & 0.0118 \\
10 & 3 & 14 &  2 & 1 & 3 & 1s(2S)2s2p(1Po) 2Po3/2 &           &  914.4365 & 7.55E+11 &      7.70E+13 & 0.0097 \\
10 & 3 & 15 &  2 & 2 & 3 & 1s(2S)2p2(1D) 2D3/2    &  920.4218 &  920.5238 & 4.25E+12 &      1.29E+14 & 0.0319 \\
10 & 3 & 16 &  2 & 2 & 5 & 1s(2S)2p2(1D) 2D5/2    &  920.4218 &  920.5326 & 4.22E+12 &      1.30E+14 & 0.0314 \\
10 & 3 & 17 &  2 & 1 & 1 & 1s(2S)2p2(3P) 2P1/2    &           &  922.5025 & 1.32E+13 &      2.00E+09 & 0.9998 \\
10 & 3 & 18 &  2 & 1 & 3 & 1s(2S)2p2(3P) 2P3/2    &           &  922.6991 & 1.32E+13 &      4.05E+11 & 0.9702 \\
10 & 3 & 19 &  2 & 0 & 1 & 1s(2S)2p2(1S) 2S1/2    &           &  932.4598 & 3.95E+12 &      1.92E+13 & 0.1706 \\
\enddata
\tablecomments{The complete version of this table is in the
electronic edition of the Journal.  The printed edition contains
only a sample.}
\end{deluxetable}


\clearpage
\begin{deluxetable}{rrrrrrr}
\tablecolumns{7}
\tablewidth{0pc} \tablecaption{\label{trans} K-vacancy transitions in the Ne,
            Mg, Si, S, Ar, and Ca isonuclear sequences}
\tablehead{\colhead{$Z$} & \colhead{$N$} & \colhead{$k$} & \colhead{$i$} &
              \colhead{$\lambda$} & \colhead{$A(k,i)$} & \colhead{$gf(i,k)$}\\
\colhead{} & \colhead{} & \colhead{} & \colhead{} & \colhead{0.1 nm} &
              \colhead{s$^{-1}$} & \colhead{}\\}
\startdata
10 & 2 &  4 & 1 & 13.5582 & 3.67E+09 & 3.04E$-$04 \\
10 & 2 &  7 & 1 & 13.4470 & 9.87E+12 & 8.04E$-$01 \\
10 & 3 &  4 & 2 & 14.1641 & 1.80E+11 & 1.09E$-$02 \\
10 & 3 &  4 & 3 & 14.1673 & 3.49E+11 & 2.10E$-$02 \\
10 & 3 &  5 & 1 & 13.8471 & 2.69E+08 & 1.54E$-$05 \\
10 & 3 &  6 & 1 & 13.8456 & 6.86E+08 & 7.88E$-$05 \\
10 & 3 &  8 & 1 & 13.6548 & 7.98E+12 & 4.46E$-$01 \\
10 & 3 &  9 & 1 & 13.6531 & 8.13E+12 & 9.09E$-$01 \\
10 & 3 & 10 & 2 & 13.8265 & 5.31E+08 & 3.05E$-$05 \\
10 & 3 & 10 & 3 & 13.8297 & 2.22E+07 & 1.28E$-$06 \\
10 & 3 & 11 & 2 & 13.8251 & 5.52E+06 & 6.34E$-$07 \\
10 & 3 & 11 & 3 & 13.8282 & 7.60E+08 & 8.72E$-$05 \\
10 & 3 & 12 & 3 & 13.8258 & 1.15E+09 & 1.99E$-$04 \\
10 & 3 & 13 & 1 & 13.5566 & 9.00E+11 & 4.96E$-$02 \\
10 & 3 & 14 & 1 & 13.5554 & 7.49E+11 & 8.26E$-$02 \\
10 & 3 & 15 & 2 & 13.7020 & 3.82E+12 & 4.31E$-$01 \\
10 & 3 & 15 & 3 & 13.7051 & 4.25E+11 & 4.79E$-$02 \\
10 & 3 & 16 & 3 & 13.7050 & 4.22E+12 & 7.13E$-$01 \\
10 & 3 & 17 & 2 & 13.6721 & 8.92E+12 & 5.00E$-$01 \\
10 & 3 & 17 & 3 & 13.6752 & 4.32E+12 & 2.42E$-$01 \\
10 & 3 & 18 & 2 & 13.6692 & 1.90E+12 & 2.13E$-$01 \\
10 & 3 & 18 & 3 & 13.6722 & 1.13E+13 & 1.27E$+$00 \\
10 & 3 & 19 & 2 & 13.5236 & 1.21E+12 & 6.67E$-$02 \\
10 & 3 & 19 & 3 & 13.5266 & 2.73E+12 & 1.50E$-$01 \\
\enddata
\tablecomments{The complete version of this table is in the
electronic edition of the Journal.  The printed edition contains
only a sample.}
\end{deluxetable}


\clearpage


\clearpage


\clearpage

\clearpage


\clearpage

\end{document}